\documentclass[journal]{IEEEtran}

\usepackage{cite}
\usepackage{amsmath,amssymb,amsfonts}
\usepackage{graphicx}
\usepackage{textcomp}
\usepackage{booktabs}
\usepackage{array}
\usepackage{tabularx}
\usepackage{url}
\usepackage[T1]{fontenc}
\usepackage[utf8]{inputenc}

\graphicspath{{figures/}}
\newcolumntype{L}[1]{>{\raggedright\arraybackslash}p{#1}}
\newcolumntype{Y}{>{\raggedright\arraybackslash}X}

\begin{document}

\title{Sensor-Stack Limits on Contactless In-Bed Body Position: A 20-Subject Multimodal Radar~+~Thermal LOSO Characterization}

\author{D\=ovy~Paukstys%
\thanks{D. Paukstys is with Komori Care, LLC, Waterford, Virginia, USA (e-mail: dovy@komoricare.com).}%
\thanks{This work was supported by Komori Care, LLC operating revenue. No external sponsor.}%
\thanks{Version 2, October 2026. This version corrects errors in version 1; see Section~\ref{sec:changes}.}%
}

\markboth{Komori Care, LLC}{Paukstys: Sensor-Stack Limits on Contactless In-Bed Body Position}

\maketitle

\begin{abstract}
Contactless in-bed body-position inference can be limited by the sensor representation the host receives rather than by the classifier. We characterize a bedside 60~GHz frequency-modulated continuous-wave (FMCW) radar that reports only on-device constant-false-alarm-rate (CFAR) point clouds, together with a low-resolution thermal array ($24\times32$ nominal, $24\times16$ effective), on 20 subjects under leave-one-subject-out (LOSO) evaluation. Two tasks share the cohort: 273 operator-directed in-bed posture holds (148.7 minutes) and a per-second bed-presence timeline (8{,}817 scored seconds). The cohort is a friends-and-family convenience sample (13 minors, ages 5 to 68, six families), and every session used the same bed in one room with the subject uncovered, so the figures describe this setup rather than a population. Bed presence reaches a 0.871 per-subject median balanced accuracy (95\% CI 0.796 to 0.910). For four-class posture, nested LOSO, in which every selection and fitting step sees training subjects only, gives 0.646 balanced accuracy (per-subject median 0.683). Prone recall is 0.45 at a precision of 0.42, so prone is not usefully detected. Version~1 reported 0.674 for a stacked ensemble picked on the folds it was scored on; this version replaces that figure with a nested estimate and corrects the bed-presence accounting and the description of the recording setting. Thermal resolves left versus right (lateral swaps of 35 to 42\% with radar alone, 8\% with thermal alone). The supine-versus-prone breathing cue shows up in the CFAR output only as a group shift (Cohen's $d=0.61$), with a clean per-hold peak in 8.4\% of holds. The next hardware test is raw range-FFT access, not further classifier tuning on CFAR detections.
\end{abstract}


\section{Introduction}
\IEEEPARstart{B}{ody} position during sleep matters clinically. The prone position is associated with elevated risk of sudden unexpected death in epilepsy (SUDEP) in several retrospective and case-control studies \cite{liebenthal2015,tao2015}. Infant safe-sleep guidance rests on placing the baby on its back \cite{moon2022}, and body position also changes obstructive sleep apnea severity \cite{cartwright1984}. More generally, in-bed posture is a confound or a target for many nocturnal-monitoring use cases. No infants took part in this study; the youngest subject was 5.

Contactless approaches to bedside posture have a mixed record, and most of the strong results come from sensors that see more of the body than ours does. Table~\ref{tab:related} lays out the adjacent work. Radar posture systems that work well read raw radio signals: raw RF reflections \cite{bodycompass2020,rfpose2018}, range-FFT features from a 60~GHz chip \cite{li2024}, or raw ultra-wideband (UWB) impulse-radar frames \cite{piriyajitakonkij2021,lai2023}. Cameras (depth, long-wave infrared, or RGB) and pressure mats see body shape or contact directly \cite{grimm2016,liu2019,liu2023,pouyan2017}, at the cost of a camera in the bedroom or a sensor under every sleeper. FMCW radar and Wi-Fi channel-state information can also track breathing and heart rate \cite{wang2020,alizadeh2019,szmola2026,liu2015}, but vital-sign accuracy is a different question from posture. Our stack sits at the sparse end: the radar firmware hands the host CFAR detections, not range bins, and the thermal array is $24\times16$ effective pixels. On CFAR-aggregated point clouds, lateral discrimination degrades because the body's centerline lies near the radar boresight (the antenna's central line of sight), so left and right map to nearly mirror-image reflection geometries. We quantify this in Section~\ref{sec:results}.

The question this paper answers is at the sensor-stack level, not the classifier level: on an off-the-shelf contactless stack made of a bedside 60~GHz FMCW radar with on-device CFAR detection plus a low-resolution thermal array, how well can in-bed body position be told apart under the tested protocol, and which sensor-layer effects bound that result? We answer empirically under leave-one-subject-out (LOSO) cross-validation across 20 subjects and name the hardware change the data make worth testing next.

This paper reports three numbers. \textbf{First}, bed presence: a 0.871 LOSO per-subject median balanced accuracy on $n=20$. \textbf{Second}, four-class posture: 0.646 balanced accuracy under nested LOSO, where the choice of pipeline is itself made inside each training split. \textbf{Third}, prone recall: 0.45, with a precision of 0.42 (19 of the 45 holds called prone are prone, Fig.~\ref{fig:confusion}), so 58\% of prone calls would be false alarms. Prone is 15.4\% of holds (42 of 273). We separate what each modality contributes, describe the mechanism behind the prone gap, and identify the hardware change worth testing.

We do not claim seizure detection in this paper. The prone-SUDEP association \cite{liebenthal2015} is motivation. Whether a contactless platform could act on it depends on whether prone can be detected reliably, and for the current hardware this paper answers that question in the negative.

\subsection{Changes From Version 1}
\label{sec:changes}
Version~1 of this preprint (June 2026) contained errors that a re-analysis in October 2026 found. Every number in this version comes from that re-analysis or was checked against it. The changes are:
\begin{enumerate}
\item \textbf{Leaky stacked ensemble.} Version~1 reported 0.674 four-class balanced accuracy for a stacked ensemble. Its meta-learner was trained on base-model probabilities from models that had already seen the test subject, and the winning pipeline was picked on the same folds it was scored on. Replaying that pipeline gives 0.660 today; the leak had little effect in the replay, and the larger problems were the selection step and tree-model drift (item 3). The headline is now nested LOSO, 0.646 (Section~\ref{sec:fourclass}). Every statistic tied to the leaky stack is withdrawn: the per-subject median 0.729 and its interval 0.675 to 0.750, the per-class recalls, the prone precision of 0.41, and the stack's lateral swap counts.
\item \textbf{Mislabeled fusion row.} The version-1 ``direct radar~+~thermal fusion'' figure of 0.602 came from gradient-boosted trees, while the surrounding text described logistic regression. Fused logistic regression scores 0.553. The tables now give every model type separately.
\item \textbf{Tree-model numbers.} Logistic regression reproduces exactly. Random-forest and gradient-boosting scores move with the random seed and library version, and several version-1 tree numbers did not reproduce. Tree columns now come from a seeded re-run and are labeled as such.
\item \textbf{Bed-presence accounting.} Version~1 summed a results file that held four approaches, so its timeline (35{,}268 seconds; 31{,}648 in bed and 3{,}620 out) and its 320 transitions were four times too large. The scored timeline is 8{,}817 seconds (7{,}912 in bed, 905 out) with 80 labeled transitions. The version-1 out-of-bed recall of 90.7\% was one minus a median false-negative rate; the median out-of-bed recall is 91.3\%.
\item \textbf{Recording setting.} Version~1 said the subjects came from eight residences and described home environments. All 20 sessions were recorded on the same bed in the same location, with fixed sensor placement and settings. Eight is the number of households the participants live in. A leave-one-residence-out split is therefore impossible; we hold out families instead, as a relatedness check.
\item \textbf{Smaller items.} The breathing-only prone recall of 0.17 did not reproduce and is now 0.21. The height correlation is recomputed on the nested per-subject scores ($r=+0.31$, was $+0.41$). The recording month is April 2026 (version~1 gave the analysis date). The latency figure, which the re-analysis did not re-check, is removed, as is the study-flow figure that carried the old numbers.
\item \textbf{Additions.} Hyperparameters and CFAR settings, a hold-duration confound check, disclosure of a non-causal thermal fill in the bed-presence features, subject-level bootstrap intervals, 30-second epoch scoring, a family holdout, and a wider related-work comparison.
\end{enumerate}
A shorter conference paper on the same data reports the corrected analysis; this preprint keeps the longer method and limitation detail.

\subsection{Contributions and Scope}
Table~\ref{tab:evaluation-ladder} gives the metric and claim boundary for each question. The contributions are:
\begin{itemize}
\item A 20-subject, subject-held-out characterization of bedside radar~+~thermal in-bed position sensing, with separate bed-presence and four-class posture tasks, reported under nested LOSO and with whole families held out.
\item A modality-level error analysis showing that thermal resolves left versus right, while the CFAR radar representation does not expose the expected supine-versus-prone breathing cue reliably per hold.
\item A scoped next-experiment claim: raw range-FFT access is the hardware lever to test next; classifier choice on host-visible CFAR point clouds did not close the prone gap in this dataset.
\end{itemize}

\begin{table*}[!t]
\caption{Positioning Against Adjacent Contactless and Low-Contact Posture Work}
\label{tab:related}
\centering
\footnotesize
\renewcommand{\arraystretch}{1.12}
\begin{tabularx}{0.97\textwidth}{L{0.16\textwidth}L{0.21\textwidth}L{0.25\textwidth}Y}
\toprule
Work or line of work & Sensor access & Evaluation shape & Relevance to this paper \\
\midrule
RF-Pose \cite{rfpose2018} & Raw radio reflections & General RF human pose estimation & Shows RF can encode body pose; says nothing about CFAR point clouds plus coarse thermal. \\
BodyCompass \cite{bodycompass2020} & Raw RF snapshots, optional per-subject calibration & Sleep posture across 26 subjects and 200+ nights; accuracy depends on how much labeled data comes from the target subject & Closest posture comparator. Supports the breathing-physics mechanism, but uses raw RF and subject labels; we test zero-label LOSO on CFAR output. \\
Li et al. \cite{li2024} & 60~GHz radar; motion statistics from range-FFT images of the raw IF samples & Supine, lateral, and prone with left and right merged; 82.74\% on 6 of 16 volunteers held out & Nearest hardware match. Its input is the range-FFT layer our firmware does not expose. \\
UWB impulse radar \cite{piriyajitakonkij2021,lai2023} & Raw UWB frames, one or two radars & Postural-transition recognition; dual-radar posture recognition & Raw radar frames carry posture and turning information. \\
FMCW vital signs \cite{wang2020,alizadeh2019,szmola2026} & Radar waveforms or range-bin access & Heart and breathing rate, including clinical settings & Radar physiology is plausible, but vital-sign accuracy is not posture accuracy. \\
Wi-Fi sleep vital signs \cite{liu2015} & Wi-Fi CSI from commodity radios & Contactless sleep vital-sign tracking & Broader contactless sleep sensing; CSI posture is outside this paper. \\
Depth, infrared, and RGB imaging \cite{grimm2016,liu2019,liu2023} & Images of the body & Sleep-position classification; in-bed pose under a blanket; a 109-participant multimodal lying-pose dataset & Sees body shape directly, but puts a camera in the bedroom. \\
Pressure mats \cite{pouyan2017} & Contact pressure under the sleeper & Posture and subject analytics from pressure maps & A strong option where a mat can be installed under each sleeper. \\
This paper & Host-visible CFAR point clouds plus effective $24\times16$ thermal frames & 20 subjects, LOSO, nested LOSO, family holdout; 273 holds and 8{,}817 scored seconds & Bed presence holds up; four-class posture is limited; prone recall is 0.45. \\
\bottomrule
\end{tabularx}
\end{table*}

\section{Method}

\subsection{Hardware and Recording Setup}
\label{sec:hardware}
The recording station consisted of:
\begin{itemize}
\item A bedside 60~GHz FMCW radar module (Texas Instruments IWR6843AOP, ES2.0) mounted at the head of the bed, running at a $\sim$10~Hz frame rate with on-device CFAR detection (range: CFAR-CASO, cell-averaging smallest-of; Doppler: CFAR-CA, cell-averaging) that produces per-frame point clouds (range, velocity, signal-to-noise ratio, azimuth, elevation). Acquisition used the \texttt{profile\_2d} configuration with two transmitters time-division multiplexed: start frequency 60~GHz; a 3.98~GHz total chirp ramp (slope 166~MHz/$\mu$s over a 24~$\mu$s ramp), giving $\sim$3.4~GHz of usable sweep bandwidth and $\sim$0.044~m range resolution; 256 ADC samples per chirp at 12.5~Msps; a 31~$\mu$s chirp (7~$\mu$s idle plus 24~$\mu$s ramp); and 64 chirps per frame (32 Doppler chirps per transmitter) at a 100~ms frame period. The antenna-on-package array ran 2 transmitters $\times$ 4 receivers (8 virtual channels). The velocity field of view is $\pm$20.16~m/s (Doppler resolution $\sim$1.26~m/s), and detections are kept between 0.25 and 9.0~m in range. CFAR-CASO in range used 8 training and 4 guard cells; CFAR-CA in Doppler used 4 training and 2 guard cells; both used a 15.0~dB threshold scale rather than an explicit probability of false alarm, and peak grouping was off in both dimensions.
\item A low-resolution thermal array (Melexis MLX90640-D110, nominally $24\times32$ pixels, rows~$\times$~columns) mounted above the bed looking down at the mattress. The D110 variant carries the wide 110\textdegree{}\,$\times$\,75\textdegree{} optic and was read at a 2~Hz full-frame refresh (the sensor streams its two subpages at 4~Hz; emissivity 0.95). The analyzed cache holds raw unmerged subpage frames, so the effective resolution is $24\times16$ (Section~\ref{sec:features}).
\item An operator-driven event tagger logging protocol-labeled pose holds (timestamped position labels with start time and duration).
\item An inertial reference: a chest-mounted M5StickC~Plus2 inertial measurement unit giving an independent gravity-axis position estimate, used as the authority for left-lateral versus right-lateral labels. The operator tag is the authority for supine versus prone. The chest unit is worn during recording and sits over the chest wall the radar images. We did not run a worn-versus-unworn radar control, so a small perturbation of the chest-wall return cannot be ruled out, and the breathing-cue analysis should be read with that caveat.
\end{itemize}

Radar and thermal streams were synchronized at the per-second tick. All analysis runs on 1~Hz feature vectors aggregated from the radar point cloud and the thermal frame. Table~\ref{tab:hardware-protocol} lists each hardware and protocol element and why it bears on the result.

\begin{table*}[!t]
\caption{Hardware and Protocol Elements Needed to Interpret the Results}
\label{tab:hardware-protocol}
\centering
\footnotesize
\renewcommand{\arraystretch}{1.12}
\begin{tabularx}{0.97\textwidth}{L{0.14\textwidth}L{0.36\textwidth}Y}
\toprule
Element & Study value & Why it matters \\
\midrule
Radar & TI IWR6843AOP, 60~GHz FMCW, head-of-bed placement, $\sim$10~Hz, host-visible CFAR point clouds & The host sees sparse detections, not raw range-FFT or raw RF. That representation is the suspected supine-versus-prone bottleneck. \\
Thermal & MLX90640, nominal $24\times32$, effective $24\times16$ in cached frames & Thermal encodes lateral bed position well, but the effective resolution is too low to resolve face versus back of head. \\
Reference labels & Operator tag for supine and prone; inertial unit for left and right & Separates visual posture labels from side-of-bed ambiguity and sets the label-noise floor. \\
Setting & One bed in one room; fixed sensor placement and settings; no blanket or sheet on any hold; operator-directed holds & Results say nothing about other rooms, beds, mounting positions, bedding, or natural sleep. \\
Task split & Bed presence from a per-second session timeline; four-class posture from analyzable in-bed holds & Keeps the easier occupancy task apart from the harder posture task. \\
Validation & LOSO across 20 subjects; nested LOSO for pipeline choice; six families held out as a relatedness check & Tests transfer to unseen people with no labeled data from them. \\
\bottomrule
\end{tabularx}
\end{table*}

\subsection{Cohort and Setting}
\label{sec:cohort}
The cohort is \textbf{20 distinct subjects} from a friends-and-family sensor-calibration group recruited by the author. Ages span 5 to 68 years, and 13 of the 20 are minors (ages 5 to 17). The subjects belong to 8 households and six families (relatives who share a family name). Household here counts where people live; it says nothing about where they were recorded.

All 20 sessions took place in April 2026 on the same bed in the same location, with the radar and thermal sensors in the same mounting positions and with the same settings every time. No subject was covered: the session log records no blanket, sheet, or comforter on any of the 331 holds. Nineteen sessions followed the full 17-pose collection script and one followed a shorter 8-pose calibration script. Each subject contributes one session, the newest available. Subject standing height was self-reported on the intake form (not measured with a stadiometer) and is used only in Section~\ref{sec:bodysize}.

\textbf{Dataset accounting.} Of 331 raw pose holds, 311 were in bed (20 out-of-bed segments are excluded from the posture task), and 273 remained after settling-time and minimum-duration filters (38 dropped for short duration or sensor-sync settling failures). The 273 holds total 148.7 minutes and split into supine 95, right lateral 74, left lateral 62, and prone 42 (prone prior 15.4\%). Table~\ref{tab:dataset-accounting} collects the hold-level and second-level counts.

\begin{table}[!t]
\caption{Dataset Accounting}
\label{tab:dataset-accounting}
\centering
\footnotesize
\renewcommand{\arraystretch}{1.1}
\begin{tabular}{@{}p{0.62\columnwidth}r@{}}
\toprule
Quantity & Count \\
\midrule
Distinct subjects & 20 \\
Households / families & 8 / 6 \\
Recording rooms (and beds) & 1 \\
Raw pose holds & 331 \\
Out-of-bed holds excluded from posture & 20 \\
In-bed holds inspected & 311 \\
Dropped for short duration or settling & 38 \\
Analyzable in-bed posture holds & 273 \\
Supine / right / left / prone holds & 95 / 74 / 62 / 42 \\
Bed-presence timeline seconds & 16{,}636 \\
Transition-flagged seconds (not scored) & 7{,}819 \\
Scored seconds, in bed / out of bed & 7{,}912 / 905 \\
Scored seconds per subject & 314--592 \\
Labeled entries / exits & 40 / 40 \\
\bottomrule
\end{tabular}
\end{table}

\subsection{Ethics, Consent, and Privacy}
\label{sec:ethics}
Consent followed the written consent form. Each adult signed it. For each of the 13 minors, a parent or legal guardian signed written permission; minors aged 8 to 17 also signed a child-assent section, and children under 8 took part on parental permission without signing. The materials described internal research and product-development data collection, voluntary participation, the right to stop or to ask for data deletion, no clinical-trial status, no medical advice or treatment, and possible de-identified aggregate use in FDA submissions, scientific posters, or publications.

These were sensor-calibration sessions. No institutional review board reviewed them, and no exemption determination was sought. Clinical validation, if pursued, is separate work under its own IRB-approved protocol and is not part of this paper.

A session could record more than the two modalities analyzed here: an inertial reference, Wi-Fi channel-state information (CSI), audio, ambient environmental sensors (e.g., light, temperature, humidity, CO\textsubscript{2}, barometric pressure), and video used to check position labels. This paper analyzes radar and thermal features only. Video, when captured, was stored on encrypted local media, viewed only to confirm labels, and then deleted (or deleted earlier on request); no video frame appears in any analysis artifact or publication. Audio is outside this analysis, and no audio waveform or speech content is reported or released. The retained posture artifacts hold de-identified radar, thermal, and inertial features and labels, not identifying media. Raw RF and IQ data are not part of the analyzed or released artifacts. Subjects appear only under random codes; the key linking codes to people is kept private and is not released.

\subsection{Bed-Presence Audit Dataset}
\label{sec:bedpresencedata}
The bed-presence analysis does not use the 273 posture holds. It uses a per-second timeline rebuilt from the same 20 sessions: we re-decode the raw radar and thermal streams for each full session (the cached posture features are in-bed only) and form one feature vector per second. Ground truth comes from the protocol log. Seconds inside an in-bed pose hold (supine, left lateral, right lateral, prone) are in bed; sitting, standing, and empty-bed seconds are out of bed; operator-logged settling and transition seconds are flagged and left out of the scores.

The timeline holds 16{,}636 seconds. Of these, 7{,}819 are transition-flagged, which leaves 8{,}817 scored seconds: 7{,}912 in bed and 905 out of bed. Each subject contributes 314 to 592 scored seconds, of which 19 to 129 are out of bed. The log contains 80 labeled changes between an in-bed posture and an out-of-bed state (40 entries and 40 exits); 52 of them involve an empty bed and 28 involve sitting on the bed. In-bed seconds outnumber out-of-bed seconds about 9 to 1, so we report balanced accuracy and per-class recall rather than raw accuracy.

The classifier is the L2 logistic regression of Section~\ref{sec:classifier}, fit on 12 per-second features (7 radar, 5 thermal) under the same LOSO protocol, with a 5-second majority filter on its output. One processing detail matters for deployment: gaps in the thermal features were filled forward and then backward across the whole session, so a missing second can borrow a value from later in the recording. That is acceptable for an offline audit, but a real-time device can only use past frames, so the presence result must be rechecked with a causal fill.

\subsection{Modalities Used in This Analysis}
We restrict this analysis to the \textbf{radar and thermal modalities}, which were available for all 20 subjects. The other modalities listed in Section~\ref{sec:ethics} are left for separate work. Keeping the modality set fixed keeps the LOSO evaluation internally consistent: every reported number uses the same sensors for every subject.

Per-modality availability across the 273 analyzable holds: radar features exist for 269 holds (98.5\%; 4 holds had too few point-cloud frames), and thermal features for 272 (99.6\%; 1 hold hit a frame-cache read error). Every hold has at least one modality. Per-experiment hold counts follow the modalities each experiment uses (the radar-only experiment uses 269 holds; the fused experiments use all 273, with the rare missing values filled inside each training fold).

\subsection{Feature Extraction}
\label{sec:features}
\textbf{Radar features.} From the per-frame CFAR point cloud we aggregate over the hold window: per-frame point counts (mean, 10th/50th/90th percentiles, standard deviation); range statistics (mean, percentiles, standard deviation); velocity statistics (mean, magnitude percentiles); SNR statistics (mean, percentiles, variance); spatial centroid and spread in $x/y/z$; and higher-moment SNR statistics, which rank highest by mutual information among the radar features.

\textbf{Thermal features.} From the per-frame thermal grid we aggregate hot-region bounding-box geometry (extents, width, height, percentiles); hot-pixel column and row centroids; the left/right hot-pixel ratio; body-centroid coordinates; coverage fraction; and percentile temperatures. The most predictive thermal category is lateral position on the bed: every top-ranked thermal feature measures where across the mattress the warm region sits.

\textbf{Cross-modal features.} Joint statistics over the two streams: the thermal column centroid normalized against the radar lateral centroid, radar lateral spread against thermal box width, and per-second co-presence flags.

\textbf{Feature sets.} The experiments use six sets, labeled E1 to E6: radar only (E1, 185 features), thermal only (E2, 355), a compact fused set (E3, 85: 39 radar, 27 thermal, 19 cross-modal), breathing features from the CFAR stream alone (E4, 82), an extended thermal set (E5, 207; Section~\ref{sec:pronegap}), and the full pool of every family (E6, 892 after the cleaning step in Section~\ref{sec:classifier}). Each posture vector aggregates the trimmed hold (mean trimmed length $\sim$27~s after a 5~s start and 1~s end trim); bed-presence vectors are per second.

The breathing band is 0.1 to 0.5~Hz (6 to 30 breaths/min). Per-hold breathing amplitude is the spectral power in that band of four per-frame point-cloud series: the SNR-weighted range centroid, the SNR-weighted $z$ centroid, the signed mean velocity, and the point count. A hold has a clean peak when the maximum in-band power is at least 2.0 times the mean in-band power.

\textbf{Three feature-pipeline limitations we disclose up front.}
\begin{enumerate}
\item \textbf{MLX90640 subpage checkerboard.} The cached frames are unmerged subpages: every other pixel sits near a 22\,\textdegree C cold reference rather than the scene temperature. That leaves 16 usable pixels in each 32-pixel row, so effective resolution is $24\times16$. A face covers 1 to 3 pixels at this resolution. Re-caching with merged subpages is a firmware fix outside this paper.
\item \textbf{CFAR radar points are sparse.} The on-device CFAR returns at most a few dozen points per frame. The chest-direct breathing-amplitude profile behind the BodyCompass mechanism \cite{bodycompass2020} needs range-resolved RF, not host-visible detections. Only 8.4\% of holds (23 of 273) show a clean breathing peak in the CFAR stream (Section~\ref{sec:pronegap}).
\item \textbf{Hold length differs by class.} The protocol held each pose for a different time: median 54~s for prone, 24~s for supine, and 14~s for each lateral side. Frame-count and segment-yield columns in E2, E3, E4, and E6 (nine of them in E6) partly encode that length. A classifier on those columns alone reaches 0.31 to 0.33 balanced accuracy, a little above the 0.25 chance level. Removing them moves the results in both directions: thermal-only logistic regression drops from 0.591 to 0.557, full-pool logistic regression rises from 0.641 to 0.662, and the full-pool random forest goes from 0.664 to 0.657. The reported numbers keep these columns. Because the length cue points toward prone, it cannot explain why prone is the weakest class.
\end{enumerate}

\subsection{Classifier and Cross-Validation}
\label{sec:classifier}
The primary classifier is \textbf{logistic regression}: L2-regularized, inverse regularization strength $C=1.0$, \texttt{lbfgs} solver, \texttt{class\_weight=balanced}, 2{,}000 to 5{,}000 iterations, with standard scaling and median imputation fit on the training subjects of each fold (E2 and E5 fill missing thermal values with zero instead). We also ran random forests (300 trees, 500 for E6) and gradient-boosted trees (\texttt{HistGradientBoosting}, 300 iterations at learning rate 0.05; E3 used 400 iterations at 0.06 with depth 6), both with balanced class weights and a fixed seed (42; 0 for E4 and E5), a random-forest variant on the top 30 features by mutual information chosen per fold, and a stacked ensemble in which a balanced logistic-regression meta-learner ($C=1$) combines the class probabilities of the logistic, forest, and boosted base models. No hyperparameter was tuned. For E6, columns that are constant or more than 30\% missing are dropped once on the pooled feature matrix before splitting; the step looks at feature values only, never labels. Logistic regression is the reference model because it is deterministic and its weights are easy to read.

Cross-validation is \textbf{leave-one-subject-out (LOSO)}: each of 20 folds fits on 19 subjects and predicts every hold of the held-out subject. Nineteen subjects contributed 14 holds and one contributed 7.

\textbf{Nested LOSO.} Version~1 picked its best pipeline on the same folds it reported, which biases the estimate upward \cite{varma2006}, and its stack fit the meta-learner on probabilities from base models trained with the test subject included. Here every outer fold runs its own inner LOSO on its 19 training subjects, which scores 20 candidates (the 19 fixed model and feature-set units plus the stack). The stack's meta-learner is fit on inner out-of-fold probabilities, so it never sees the outer test subject. The inner winner is refit on all 19 training subjects and applied to the held-out subject. The four-class headline is the aggregate balanced accuracy of these outer predictions.

\textbf{Family holdout.} Relatives tend to share build, which could flatter subject-level splits. As a relatedness check we also hold out each of the six families in turn. All sessions shared one room and one bed, so a split by residence is not possible with this cohort.

\textbf{Statistics.} We report aggregate balanced accuracy (from the pooled confusion matrix of all held-out predictions) and the per-subject median with its interquartile range. Confidence intervals on medians come from 10{,}000 bootstrap resamples of subjects (seed 20261003), so correlated seconds or holds within a subject are never treated as independent. Chance balanced accuracy is 0.25 for four classes and 0.50 for bed presence. Table~\ref{tab:evaluation-ladder} sets out the questions, units, metrics, and claim boundaries.

\begin{table*}[!t]
\caption{Evaluation Ladder and Claim Boundary}
\label{tab:evaluation-ladder}
\centering
\footnotesize
\renewcommand{\arraystretch}{1.12}
\begin{tabularx}{0.97\textwidth}{L{0.13\textwidth}L{0.15\textwidth}L{0.16\textwidth}L{0.21\textwidth}Y}
\toprule
Question & Unit and split & Primary metric & Baselines or ablations & Claim boundary \\
\midrule
Is someone in bed? & Per-second timeline, LOSO by subject & Per-subject median balanced accuracy with subject-bootstrap CI & Fused L2 logistic regression; 30~s epochs; family holdout; chance 0.50 & Steady-state protocol audit with transitions removed; not an overnight false-alarm rate. \\
Can the stack classify four postures? & Per-hold features, nested LOSO by subject & Aggregate balanced accuracy, per-subject median and IQR & E1 to E6 under logistic regression, forest, boosting, and stacking; chance 0.25 & Nested 0.646 is the estimate; fixed-pipeline rows are references, not a ceiling. \\
Which modality matters? & Per-class errors under LOSO & Lateral swap rates and prone recall & Radar only, thermal only, fused, full pool & Thermal earns its place on left versus right; radar adds no prone cue in the CFAR feature path. \\
What blocks prone? & Prone row and CFAR breathing checks & Recall, precision, clean-peak rate, Cohen's $d$ & E1 to E6, breathing-only E4 & Prone recall 0.45 nested, at most 0.49 in any LOSO row; raw range-FFT is a next experiment, not a shown fix. \\
\bottomrule
\end{tabularx}
\end{table*}

\subsection{Ground-Truth Labels}
\label{sec:labels}
Position labels follow a hybrid scheme:
\begin{itemize}
\item \textbf{The operator tag decides} supine and prone (the operator checks the posture at hold start).
\item \textbf{The inertial reading decides} left lateral versus right lateral (the operator's side-of-bed view is unreliable; the gravity axis on the chest unit is reliable).
\item Sitting, standing, and out-of-bed holds are labeled by operator tag; they supply out-of-bed seconds for the presence timeline and are excluded from the four-class task.
\item Each hold loses its first 5~s and last 1~s, which removes transition and settling frames.
\end{itemize}
The inertial unit reported a position label about every 10~s (raw 10~Hz samples were not logged for these sessions). Left or right was assigned from the dominant gravity component (left lateral when the device $+X$ axis points up, right lateral when $-X$ points up), accepted when the gravity-projection cosine exceeded 0.65.

Of the 311 in-bed holds checked for consistency, 7 (2.3\%) showed disagreement between the operator tag and the inertial position; they were resolved by the authority rules above. This 2.3\% rate is a label-noise floor for the per-class results.

\subsection{Reproducibility}
\label{sec:reproducibility}
Feature extraction, the classifier pipeline, and the LOSO harness are version-controlled at Komori Care, with run configurations and per-subject outputs stored with each run. The October 2026 re-analysis behind this version ran on Python~3.12.9 with scikit-learn~1.9.0. Logistic-regression results are deterministic and match the original runs exactly. Tree-based results depend on the seed and library release: across seeds 0 to 9, the radar-only forest ranges from 0.448 to 0.491 and the full-pool forest from 0.663 to 0.683. The source package includes an ancillary bundle under \path{anc/paper_B_v2_release_bundle_2026-10-04/} with dataset accounting, aggregate results, an evaluation manifest, a per-subject metric schema, a claim-to-evidence matrix, a release-boundary note, and a list of corrections from version~1. It contains no participant identifiers.

Per-experiment artifacts under random subject codes (feature matrices, per-fold confusion matrices, per-subject balanced accuracies) are available from the author on reasonable request, subject to participant consent and venue policy. Raw RF and IQ data, raw point-cloud streams, raw thermal frames, video, audio, consent documents, household identifiers, and the code-to-identity key are not released. The feature-extraction and classifier code are proprietary to Komori Care and are not released.

\section{Results}
\label{sec:results}

\subsection{Bed-Presence Binary Classification}
\begin{figure}[!t]
\centering
\includegraphics[width=\columnwidth]{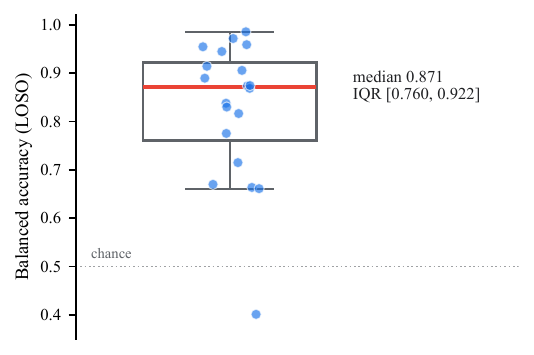}
\caption{Bed-presence balanced accuracy for each of the 20 held-out subjects (fused radar~+~thermal logistic regression, LOSO). The box spans the interquartile range with the median (0.871) as its center line; whiskers reach 1.5 times the IQR; each dot is one subject. The dotted line is chance (0.50); the one subject below it scores 0.40.}
\label{fig:bedpresence}
\end{figure}

Fused radar~+~thermal logistic regression reaches a \textbf{0.871 per-subject median LOSO balanced accuracy} for in bed versus out of bed (interquartile range 0.760 to 0.922; subject-bootstrap 95\% CI 0.796 to 0.910). Per class:
\begin{itemize}
\item \textbf{In-bed recall}: 95.2\% (median), so 4.8\% of in-bed seconds are called out of bed.
\item \textbf{Out-of-bed recall}: 91.3\% (median), so 8.7\% of out-of-bed seconds are called in bed.
\end{itemize}
These three figures are separate per-subject medians and do not combine: the mean of the two recall medians is about 0.93, above the 0.871 balanced-accuracy median, because a subject who is weak on one class loses balanced accuracy while the class medians stay high. Nineteen of 20 subjects score above chance, and one sits below it at 0.40 (Fig.~\ref{fig:bedpresence}).

The subject is the unit here because sessions differ in length and consecutive seconds are strongly correlated. Two other views give similar or somewhat lower figures. Pooling all scored seconds, so that longer sessions weigh more, gives 0.814 (in-bed recall 0.844, out-of-bed recall 0.785). Scoring non-overlapping 30-second epochs by majority vote, 332 epochs of which 35 are out of bed, gives a per-subject median of 0.896 (interquartile range 0.75 to 1.0; 95\% CI 0.804 to 1.0) and 0.851 pooled. With whole families held out, the per-subject median is 0.848 (interquartile range 0.752 to 0.901; 95\% CI 0.777 to 0.893).

Two cautions apply. Out-of-bed data are thin: 905 scored seconds and 35 epochs across 20 subjects, so the out-of-bed rate rests on little data. The thermal gap-fill described in Section~\ref{sec:bedpresencedata} is non-causal and has to be replaced before an online test.

\textbf{Interpretation.} Bed presence is the easier problem in this protocol. On this thermal~+~radar stack, in or out of bed can be told apart without contact for most subjects. This is a steady-state audit of directed holds with transitions removed. It is not a transition-timing result or an overnight false-alarm rate.

\subsection{Four-Class Position Classification}
\label{sec:fourclass}
On the same stack and the same 20 subjects, we now ask how well the 273 holds can be sorted into supine, left lateral, right lateral, and prone.

\textbf{Fixed pipelines.} Table~\ref{tab:configs} gives the LOSO balanced accuracy of each feature set under each model type. With logistic regression, radar alone reaches 0.440, thermal alone 0.591, and the compact fused set 0.553, which is lower than thermal alone. The full pool of 892 features reaches 0.641. The tree models do somewhat better on the fused and full-pool sets (0.605 and 0.664 for the forest), but those values carry the seed spread noted in Section~\ref{sec:reproducibility}. The mutual-information top-30 forest scores 0.613. None of these fixed rows is an estimate of the best pipeline's performance, because choosing the highest row after seeing the scores is itself a fitting step.

\begin{table}[!t]
\caption{Four-Class LOSO Balanced Accuracy by Feature Set and Model. LR is deterministic; RF and HGB come from the seeded re-run and vary with seed.}
\label{tab:configs}
\centering
\footnotesize
\begin{tabular}{@{}lrccc@{}}
\toprule
Feature set & Holds & LR & RF & HGB \\
\midrule
E1 radar only (185) & 269 & 0.440 & 0.448 & 0.418 \\
E2 thermal only (355) & 272 & 0.591 & 0.574 & 0.585 \\
E3 compact fused (85) & 273 & 0.553 & 0.605 & 0.601 \\
E4 CFAR breathing (82) & 273 & 0.347 & 0.397 & 0.380 \\
E5 extended thermal (207) & 271 & 0.521 & 0.565 & 0.539 \\
E6 full pool (892) & 273 & 0.641 & 0.664 & 0.660 \\
\midrule
\multicolumn{2}{@{}l}{E6 stack, meta-learner nested} & \multicolumn{3}{c}{0.664} \\
\multicolumn{2}{@{}l}{Nested LOSO with selection} & \multicolumn{3}{c}{\textbf{0.646}} \\
\bottomrule
\end{tabular}
\end{table}

\begin{figure}[!t]
\centering
\includegraphics[width=\columnwidth]{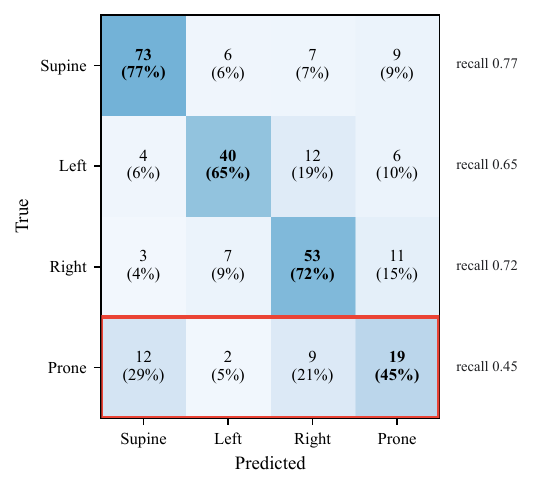}
\caption{Four-class confusion matrix for the nested LOSO predictions (0.646 balanced accuracy, 273 holds). Each cell gives the count and the row percentage; per-class recall is printed to the right. The outlined prone row is the weakest: 19 of 42 prone holds are found, and 19 of the 45 prone calls are correct.}
\label{fig:confusion}
\end{figure}

\textbf{Nested result.} With the choice of pipeline made inside each training split, the outer folds give \textbf{0.646 aggregate balanced accuracy}. The per-subject median is 0.683 (interquartile range 0.531 to 0.750; 95\% CI 0.562 to 0.746), and per-subject scores range from 0.350 to 0.950 (standard deviation 0.164). Per-class recall is 0.77 supine, 0.65 left lateral, 0.72 right lateral, and 0.45 prone; precision is 0.79, 0.73, 0.65, and 0.42. The inner loop picked the full-pool forest in 12 folds, the stack in 7, and the full-pool boosted model in 1. It never picked a radar-only, thermal-only, or compact fused unit. Fig.~\ref{fig:confusion} shows the confusion matrix.

The stack with its meta-learner nested but with no selection step scores 0.664, and so does the best candidate picked after looking at the outer folds; the honest selection step costs about 0.02. Replaying the version-1 stack, leak included, gives 0.660 rather than the 0.674 published. In this replay the leak itself made little difference, and the rest of the gap is seed and library drift in the tree models. The version-1 figure was the best of many pipelines scored on the folds it reported, which is why the nested number replaces it.

\textbf{Family holdout.} Holding out whole families leaves the ordering intact: radar only 0.356, thermal only 0.574, and full pool 0.629 with logistic regression; 0.610 for the fused boosted model; and 0.669 for the stack with its meta-learner nested (prone recall 0.50, precision 0.47). Relatedness between training and test subjects does not appear to drive the subject-level results. This check cannot speak to other rooms or beds.

\subsection{Which Modality Solves Which Sub-Problem}
Left versus right lateral is the clearest modality finding (Table~\ref{tab:swap}).

\begin{table}[!t]
\caption{Lateral Swap Rates. Denominators are the per-class holds in each configuration's set; radar-only uses the 269 holds with radar features (72 of 74 right-lateral holds).}
\label{tab:swap}
\centering
\footnotesize
\begin{tabular}{@{}lcc@{}}
\toprule
Configuration & L$\rightarrow$R & R$\rightarrow$L \\
\midrule
E1 radar only, LR & 26/62 (42\%) & 25/72 (35\%) \\
E2 thermal only, LR & 5/62 (8\%) & 6/74 (8\%) \\
E3 compact fused, LR & 9/62 (15\%) & 11/74 (15\%) \\
E6 full pool, LR & 7/62 (11\%) & 5/74 (7\%) \\
Nested LOSO with selection & 12/62 (19\%) & 7/74 (9\%) \\
\bottomrule
\end{tabular}
\end{table}

Radar alone swaps the two lateral classes 35 to 42\% of the time. Thermal alone swaps them 8\% of the time. The cause is geometric: with the radar at the headboard, the body's sagittal (head-to-toe midline) plane lies near the boresight, so a left roll and a right roll look like mirror images in the spatial-aggregate features of the CFAR point cloud. The thermal array sees the warm region move about 5 columns across the mattress as the subject rolls, and the thermal features capture that shift directly.

\textbf{Thermal is in the stack because of left versus right.} That one sub-problem is the strongest argument for keeping the thermal sensor.

Supine versus prone is harder. Both postures put the body in the middle of the mattress, both look alike at $24\times16$ thermal resolution, and both give a similar radar SNR signature. The expected differentiator is chest-direct breathing amplitude, the mechanism BodyCompass reports for raw-RF sleep posture \cite{bodycompass2020}: lying supine, the chest wall moves toward the radar; lying prone, the back moves and the chest is hidden. The CFAR stream shows a compatible group-level signal: per-hold breathing-band amplitude differs between the 95 supine and 42 prone holds with \textbf{Cohen's $d=0.61$}. But only 8.4\% of holds yield a clean per-hold peak, and the breathing-only classifier (E4) finds at most 21\% of prone holds. BodyCompass works on raw RF snapshots and reports 94\%, 87\%, and 84\% accuracy with one week, one night, and 16 minutes of labeled data from the target subject; our evaluation uses no labeled data from the held-out subject and runs on CFAR-compressed firmware output, which is a harder setting.

\textbf{The breathing cue is not reliably available per hold from CFAR point clouds.} The group shift ($d=0.61$) fits the mechanism, but a classifier needs the cue in each hold, and only 8.4\% of holds provide it.

\subsection{The Prone Gap}
\label{sec:pronegap}
Table~\ref{tab:prone} gives prone recall for every configuration. E1, E2, E3, and E6 are defined in Section~\ref{sec:features}. E4 uses only the CFAR breathing features. E5 is a thermal-only set that adds shape and orientation features the E2 set lacks: seven log-scaled Hu moments, fitted-ellipse geometry (major and minor axes, eccentricity, orientation, area), region solidity, a face-visibility flag, and bilateral-symmetry features about the body-centroid column.

\begin{table}[!t]
\caption{Prone Recall by Configuration (42 Prone Holds; 41 in the Thermal-Only Sets). RF and HGB Are Seeded Re-Runs.}
\label{tab:prone}
\centering
\footnotesize
\begin{tabular}{@{}lccc@{}}
\toprule
Configuration & LR & RF & HGB \\
\midrule
E1 radar only & 0.39 & 0.34 & 0.20 \\
E2 thermal only & \textbf{0.49} & 0.41 & 0.41 \\
E3 compact fused & 0.36 & 0.19 & 0.31 \\
E4 CFAR breathing only & 0.21 & 0.07 & 0.19 \\
E5 extended thermal & 0.37 & 0.41 & 0.29 \\
E6 full pool & 0.48 & 0.45 & 0.29 \\
\midrule
Nested LOSO with selection & \multicolumn{3}{c}{0.45 (precision 0.42)} \\
\bottomrule
\end{tabular}
\end{table}

No LOSO configuration finds half of the prone holds. The nested result finds 19 of 42, and only 19 of its 45 prone calls are right, so 58\% of its prone alarms would be false; for an always-on monitor, the false-alarm load is as much a cost as the misses. Thermal-only logistic regression (0.49) is the best single row: radar, fusion, and the full pool add nothing to prone recall over thermal alone. The compact fused set (0.36 with logistic regression) is worse than either sensor by itself (radar 0.39, thermal 0.49), so the post-CFAR radar columns widen the feature space but carry no prone information. The one radar cue expected to separate supine from prone is breathing amplitude, and CFAR makes it unreliable per hold before the host sees it (E4, at most 0.21). Three reasons work together:
\begin{enumerate}
\item \textbf{Class imbalance.} Prone is 42 of 273 holds (15.4\%); supine is 95 (34.8\%). Balanced class weights only partly correct for this.
\item \textbf{Weak visible features.} Prone shows the back of the head to the thermal array. Hair insulates, so the head reads cooler, but at 1 to 3 pixels per face the head-orientation signature does not separate cleanly from supine.
\item \textbf{Breathing physics is visible in aggregate but unreliable per hold.} Cohen's $d=0.61$ on breathing amplitude (supine above prone, as BodyCompass predicts \cite{bodycompass2020}), yet only 8.4\% of holds give a clean peak from CFAR detections.
\end{enumerate}

We attribute the prone gap mainly to the CFAR representation, but with about 2 prone holds per subject under a fixed pose script we cannot fully separate a representation limit from thin prone sampling. A study with natural, repeated prone episodes would be needed to separate the two.

The most plausible lever for raising prone recall is \textbf{raw range-FFT access}: per-range-bin chest-motion amplitude instead of on-device CFAR detections. Multi-bin radar recovers breathing rate in clinical monitoring \cite{szmola2026}, and the strongest 60~GHz posture result we found builds on range-FFT features \cite{li2024}; that per-bin access is exactly what the current firmware path withholds. A raw-data bridge is the planned hardware step. We name it as the next test, not a proven fix: the data point to the on-device CFAR layer as the place where the breathing cue becomes unreliable for host-side features, but only a raw range-FFT experiment on the same subject-held-out protocol can show whether the cue is enough to close the prone gap. On this CFAR point-cloud stack, prone is the binding constraint, and none of the pipelines tested closed it.

\subsection{Body-Size Trend}
\label{sec:bodysize}
Version~1 reported $r=+0.41$ between subject height and per-subject balanced accuracy, computed on the leaky stack. On the nested per-subject scores the correlation is \textbf{$r=+0.31$} ($n=20$, $p=0.19$; 95\% CI $-0.16$ to $+0.66$ by Fisher $z$; Spearman $\rho=+0.32$). The interval spans zero, so this is at most a weak directional hint. Height also tracks age here, since 13 of the 20 subjects are minors, so a body-size sensing effect cannot be told apart from a child-versus-adult effect at this sample size. Two mechanisms would push in the observed direction: lateral-position features are in pixel coordinates, so a small body that does not span the field of view moves its centroid less per degree of roll, and a small body gives lower thermal contrast and a weaker radar return. Body-relative coordinates or a short install-time calibration could address both; neither is tested here. We keep the analysis because the smallest subjects are children and any pediatric use would need the trend checked on a larger sample.

\section{Discussion}
\label{sec:discussion}
\subsection{Summary of Findings}

\textbf{Bed presence holds up in this protocol.} The per-subject median is 0.871 (95\% CI 0.796 to 0.910), median per-class recall is above 90\% for both classes, and the result survives 30-second epoch scoring and family holdout. It comes from operator-directed holds on one bed in one room with no bedding and with transitions removed, and the thermal gap-fill used future frames. Within those limits, it is strong enough to justify testing a ``person in bed'' gate for downstream nocturnal monitoring. The layers such a gate would feed, such as prolonged-immobility or bed-exit flags, are future work and are not evaluated here.

\textbf{Four-class position is limited by the exposed representation.} Nested LOSO gives 0.646. Direct fusion with logistic regression is below thermal alone, and the full pool helps by a few hundredths at most, an amount within the subject-level spread at $n=20$. We read this as the limit of what the tested pipelines extract from this representation, not as evidence of strong cross-modal synergy and not as a physical ceiling.

\textbf{Prone is the binding constraint.} Prone recall is 0.45 under nested LOSO and at most 0.49 in any LOSO configuration tested, with a nested precision of 0.42. The breathing mechanism that should separate supine from prone is visible as a group shift in the CFAR point clouds ($d=0.61$) but is unreliable per hold. Raw range-FFT access is the lever to test; the raw-RF literature \cite{bodycompass2020,li2024} supports the expectation that range-resolved data carry posture information, but we have not measured that path on this hardware. Because these were stable, directed holds rather than natural sleep, a deployed system on the same CFAR stack would likely do worse.

\textbf{Thermal earns its place by solving left versus right.} With the radar at the headboard, lateral swaps run 35 to 42\% on radar features alone. Thermal brings them to 8\%. The aggregate gain from adding thermal is modest, but it lands on the one sub-problem that radar geometry cannot solve.

\textbf{Limitations.} One bed in one room, with fixed sensor placement and settings, so nothing here transfers to other rooms, beds, or mounting positions without testing. No bedding on any hold. Calibration-protocol data only: subjects held positions on instruction, not overnight sleep with natural transitions, blankets, or partial rolls. A friends-and-family convenience cohort with many children, from six families. One hardware iteration and one session per subject, so within-subject stability across nights is untested. Twenty subjects is small for four-class LOSO: the nested per-subject scores run from 0.350 to 0.950 (standard deviation 0.164). Several feature columns encode hold length, which differed by class. The bed-presence thermal fill is non-causal. The MLX90640 frames are effectively $24\times16$. Tree-model results move with the seed. The 2.3\% label disagreement rate sets a floor on per-class uncertainty.

\textbf{Forthcoming work.} First, raw range-FFT access through a firmware data bridge, to test whether range-resolved radar data improve supine-versus-prone separation on this hardware under the same subject-held-out protocol. Second, recordings with bedding, other rooms, and natural sleep. Wi-Fi CSI sensing is left for separate work.

\section{Conclusion}
We report what contactless radar~+~thermal sensing can and cannot do for in-bed body position on a CFAR point-cloud plus low-resolution thermal stack, evaluated with subjects held out across 20 subjects recorded in one room. Bed presence reaches a 0.871 per-subject median balanced accuracy. Four-class position reaches 0.646 under nested LOSO (per-subject median 0.683). Prone recall, the sub-problem with the strongest clinical motivation, is 0.45 and limits the whole configuration. Thermal solves left versus right; radar carries a plausible breathing route to supine versus prone, but the CFAR feature path exposes that cue only in aggregate, not in the single hold where classification needs it.

This is a sensor-stack characterization, not a best-system paper. What comes next is a raw range-FFT test on this same hardware, again with subjects held out.

\section*{Data and Code Availability}
The source package includes the ancillary bundle described in Section~\ref{sec:reproducibility} under \path{anc/paper_B_v2_release_bundle_2026-10-04/}. It replaces the version-1 bundle, whose counts and results carried the errors listed in Section~\ref{sec:changes}. It holds aggregate counts and results only and no participant-identifying material. De-identified per-experiment artifacts may be shared for confidential review when consistent with participant consent and venue policy. Proprietary feature-extraction and evaluation code are not released.

\section*{Acknowledgment}
This work was performed by Komori Care, LLC, Waterford, Virginia. The 20-subject cohort is friends-and-family calibration data and is not part of any IRB-reviewed protocol; consent is described in Section~\ref{sec:ethics}. We thank the participants and their families for their time and patience during the calibration sessions.

Manuscript drafting and editorial review were assisted by AI tools (OpenAI ChatGPT and Anthropic Claude) for organization, wording, critique, and polishing. The October 2026 re-analysis code was written and run with AI coding assistance (Anthropic Claude Code) under the author's direction. The tools did not generate sensor data, assign participant labels, or determine final claims. The author is solely responsible for all data, methods, claims, and conclusions presented.

\end{document}